\documentstyle[12pt]{article}
\begin{document}

\newcommand{\JWN}{J.W.~Norbury}
\newcommand{\FAC}{F.A.~Cucinotta}
\newcommand{\APJS}{Astrophysical Journal Supplement}
\newcommand{\TM}{NASA Technical Memorandum}

\vspace{.05in}
\begin{center}
{\Large Revitalization of an undergraduate physics program }\\
\vspace{.2in}
{\large John W. Norbury}\\
{\em Physics Department,University of Wisconsin-Milwaukee, 
P.O. Box 413, Milwaukee, Wisconsin 53201}, e-mail: norbury@csd.uwm.edu\\
\vspace{.1in}
{\large G. R. Sudhakaran}\\
{\em Physics Department, University of Wisconsin-La Crosse, 
La Crosse, Wisconsin 54601}, e-mail: sudhak@physics.uwlax.edu\\

\vspace{.1in}
\end{center}

\noindent
This article describes the  successful revitalization of an
 undergraduate physics program. The areas of curriculum development, undergraduate research experiences
and advising and retention, to name a few, are emphasized in this interconnecting and systematic approach
whereby each and every effort combines to get results.  The program   can be used by other
physics departments wishing to improve and expand undergraduate education in physics.

\section{Introduction}

Undergraduate physics programs in the United States and other countries seem
 to be in a period of decline. There are less students taking the physics major  and less students entering
graduate study [1]. Some departments are being closed down and others are under the threat of
closure.  In such circumstances demoralization sets in and spirals into every aspect of a program
discouraging faculty and students.  

Five years ago  the physics department at the 
University of Wisconsin - La Crosse had a total of 5 physics majors, 5 faculty and a
graduation rate of about one physics major every two years. The department had received a poor review and
was in danger of being phased out, but instead of taking this easy option the dean of the college decided
instead to hire a new chair in an attempt to turn the department around.  Five years later the department is
one of the best on campus, has received an excellent review and currently has a total of about 85 physics
majors and 7 faculty. The present article describes how this was achieved. It is hoped that the information
presented here can be used by other physics departments to revitalize their programs.

\section{Program outline}

\underline  {\it 1. Academic Programs.}
The first thing was to change the academic programs being offered
 and re-package them in attractive ways directing students, parents and teachers to expand their typical
view of what a physics degree could do for the student.  We still continued the core subjects of modern
physics, mechanics, electrodynamics, quantum mechanics, thermodynamics and optics. We  also continued two
popular astronomy courses and the introductory year long sequences of  algebra and calculus based physics
courses. However several new courses were added to make the elective list a lot more interesting and useful
for the students. Some of the electives added were   quantum optics, electronics, seminar (for
credit), research (for credit), computational physics and advanced computational physics, general
relativity and cosmology, astrophysics, advanced quantum
mechanics and particle physics.  

\underline  {\it 2. Emphases and Concentrations.}
One of the important additions in attracting new majors was the
 introduction of a set of emphasis programs that could be packaged along with course and career information.
These included physics major with business concentration, physics major with astronomy emphasis, physics
major with computational physics emphasis and  physics major with optics emphasis.

The physics major with business concentration basically consisted 
of a physics major with a business minor for a total of about 55 credits. Why go to the trouble of simply
re-packaging an already existing product (physics major and business minor) into a 'new' product (physics
major with business concentration)? This is an important point and should not be lost. Physics now needs to
be marketed as does any other product, and the department has to have the products to suggest and then
deliver to students.  Physics programs now need to be attractive not only to students, but also to parents
and teachers, who heavily influence the students. Often these clients simply don't think of physics and
business as going together, yet most of us in the field know that this is an excellent combination for
students wishing to obtain employment with a bachelors degree. Having a formal program such as a physics
major with business concentration highlights the career opportunities available to students and simply makes
the overall physics program look a better match in today's job market. A quick perusal of job sections in
newspapers show marketing/sales in technical areas, computer and technology skills being needed in the
business sector. These are all skills taught in a physics degree (with business concentration).

 In order to be able to offer these emphases the department had to 
add quite a few courses to the catalog, as mentioned above. The areas of optics, computational physics and
astronomy were chosen deliberately. Optics is very important for industry and is a good area for job
seekers. Computational physics  is also an excellent area for both job seekers and those wishing to go to
graduate school. Astronomy was chosen simply because so many students have an interest in this area. It was
also important to make sure that the department had expertise in these areas. Aside from offering the
regular physics major, three areas of emphasis were introduced namely optics, astronomy and computational
physics. The total number of credits for these three programs was similar to the regular physics major, but
the elective options were eliminated. Instead, the electives were chosen for the students. For the astronomy
emphasis the student was required to take the core physics courses plus 3 astronomy courses and a
research project in astronomy. Similarly for the optics and computational physics emphasis. An 
optics experimentalist  was hired to help with this. One can easily imagine other
departments with different areas of expertise developing different emphasis programs. There are  many
departments that already have an extensive listing of electives. From the student, parent, teacher point of
view however, the existence of these electives and what they can do is often lost. It is very worthwhile to
package some strong elective programs into emphasis areas so that it is clear as to what specializations are
available to students and how this relates to the real world task of getting a job. 

\underline  {\it 3. Honors Program.}
A physics honors program was also introduced, in which students are 
required to submit a formal application, maintain a certain GPA,  complete a research project with
distinguished performance, give a seminar and be recommended by two faculty members.

\underline  {\it 4. Dual degree in physics and engineering.}
One of the most important programs introduced was a dual degree program
 in physics and engineering. Such programs are starting to gain popularity and are an excellent way to
revitalize an undergraduate physics department. The program introduced was a collaborative program between
our own department and two engineering schools (University of Wisconsin - Madison and University of
Wisconsin - Milwaukee).  An essential feature of this program is that a student is guaranteed acceptance
into the engineering school upon completion of a set of physics and other required courses with a specified
grade point average. The students spend 3 years in the department at the University of Wisconsin - La Crosse
studying selected physics courses and then transfer to one of the engineering schools for 2 years to study
an area of engineering. After the first year at the engineering school the student receives a physics degree
from La Crosse and after the second year receives an engineering degree, thus graduating with two degrees
that complement each other.   This program has been extremely attractive to students, parents and teachers.
We strongly recommend such a program to undergraduate physics departments. 

\underline  {\it 5. Laboratory Upgrades.}
As part of improving the academic programs, a lot of attention was paid 
to upgrading the laboratory facilities. During the past five years approximately \$200,000 in laboratory
modernization funds were spent in upgrades. One cannot expect students first and facilities later. They
only come together. The freshman physics labs were completely overhauled using computer based "workshop
physics" style laboratories. The students went from hating lab work to actually enjoying it.  In
addition the modern physics lab, optics lab and electronics lab were completely re-done with a full
complement of modern experiments and equipment.

\underline  {\it 6. Quality Instruction.}
The quality of instruction in all courses (but especially the 
introductory courses) was improved by trying very hard to use the best instructors. This seems mundane, but
is extremely important in building up a physics program. If the majors have a couple of, or even one,  poor
instructor then the program suffers tremendously. It is absolutely vital to have high quality instruction
and every effort must be made to make sure this happens.

\underline  {\it 7. Undergraduate Research.}
One of the major factors that lead to high student satisfaction with our new program was a strong set of 
research experiences for the undergraduate physics majors. Before we came to the department, research was
almost non-existent. As an incentive to faculty  the chair allowed supervision of undergraduate research to
count for one course for the faculty member and as an incentive for students we introduced research for
credit that a student could take. Further to this a research experience was a requirement for each of the 3
emphasis programs described above. Three out of the six faculty became actively involved in  student
research projects immediately. Research was offered in all three areas of experiment, theory and computation
in the areas of optics, condensed matter physics, particle physics, nuclear physics and astronomy.  Several
students worked only on semester-long projects, but the most successful experiences were with students who
would work for two semesters and one summer. The advantage of this work is many-fold. Fliers to schools,
brochures, campus news, department annual reports, student  and faculty resumes all are enhanced.

Many good undergraduate departments have such an undergraduate research
 program in place. When talking to administrators about physics we often emphasize the triad of theory,
experiment and computation. When talking about physics education we emphasize the triad of lecture,
laboratory and research. It is vitally important for undergraduates to have a good research experience
during their education. It also helps the department atmosphere tremendously and is very good to display
when giving tours to students, parents, teachers and administrators.

\underline  {\it 8. Student Presentations.}
The research outlined above could be show-cased to other students or
 advertised, and so attracted further students. Students were encouraged  and trained to present the results
of their work at department seminars and at conferences such as the
Argonne symposium [2-9] and also at national and international meetings such as the American
Astronomical Society and the International Symposium on Molecular Spectroscopy [10-13]. Many
students and faculty published papers together [14-21]. Faculty benefitted also by being able to place
any of the work related to student research  in their promotion files.

\underline  {\it 9. Funding for Students.}
Funding was obtained   so that students could work on research over
 the summer. This also gave the department the opportunity to give students and parents the promise of
monetary support and see the immediate connection between learning physics and monetary gain. Again any
student getting such support was used for real promotional advantages in the department literature and
annual reports.

\underline  {\it 10. Scholarships and Internships.}
We went to great efforts to have the students apply for scholarships and internships. Several students won
very prestigous scholarships (e.g Barry Goldwater scholarship, Council on Undergraduate Research
Felllowship, American Physical Society Summer Fellowship) and this had a strong effect on the motivations of
the other students. Summer interns were also arranged. One of the best programs here is the `Research
Experiences for Undergraduates' run by the National Science Foundation.  Again a great deal of busy work is
involved in arranging scholarships and interships but the work is certainly worthwhile. It also helps a lot
with recruitment in being able to give examples of the successes of previous students.

\underline  {\it 11. Seminar Program for Credit.}
There are several other  elements that went into building up the physics
 program. One was the establishment of a department seminar program. This was specifically designed to
provide a meeting place for the majors and faculty.  We introduced a program where students could sign up for
1 credit of course work. The requirement was to attend all the seminars and to either write a report on one
of them or present a seminar. What was interesting about this was that many students outside of physics also
signed up. Many physics majors did not sign up but attended anyway and the group grew.  Speakers included
faculty from physics and other departments, physics majors and outside speakers. The physics majors would
often talk about their research projects and this was a great way for other students to see what
opportunities were available. Students also talked about their summer internship experiences.  Outside
speakers gave talks primarily on research topics, but there were also talks on careers and engineering
programs. 

\underline  {\it 12. Recruitment, Advising, Retention.}
Recruitment and advising appears at first to be another area that seems 
to be very mundane. However our experience is that the role of the undergraduate physics major advisor is
absolutely essential for a successful physics program. The advisor should be very knowledgeable about
the employment situation,  salaries, current job openings, scholarships, internships, summer jobs, tutoring
jobs, housing, international opportunities, graduate record exam, graduate schools, etc.The physics advisor
needs to be constantly available, always happy and willing to spend lots of time with the students, have a
friendly personality, and do many other tasks too numerous to mention. Not all faculty find all this busy
work palatable, but in our experience it is one of the most important factors in {\em retaining} students
once they sign up for the major.

\underline  {\it 13. Advertising and Brochures.}
Advertising is another extremely important area that needed attention. 
One can have the best physics program in the world, but if no one knows about it, then not much is going to
happen. The primary way of advertising was to be in touch with physics high school teachers and counselors
and to let them know of the new programs that were available with regular mail outs.  Teachers were sent
information about actual student work as well as general programs so they could give this  immediately to
their own students.  Teachers were also invited regularly to the department seminars and social gatherings.
We believe that letting teachers know about the unique aspects of a physics program is one of the best ways
to bring in new majors. 

\underline  {\it 14. Presenting  a Plan and Cooperating with Administration.}
Another  aspect of building up the physics program was cooperation and 
interaction with the university administration. This included not only the deans, provost and chancellor,
but also people in the international office, the career center, the counseling center, the affirmative
action office, the library,  the computer center, etc. It is vitally important to have a good relationship
with all of these areas and to explain your plan and future directions. The dean, vice chancellor and
chancellor were especially important. When building up a program it is essential to obtain financial
commitments and to have these commitments followed through. Often these groups were invited to the
department seminars or demonstrations or we provided a tour for administration visitors.

\underline  {\it 15. Department Team Work and Priority Mission.}
Finally we should mention the obvious, that all of the above {\em cannot} be 
done by one person as every aspect needs attention. No one idea is a quick fix that will work but  
a sustained concerted effort is needed  over several years. We were very fortunate to have a
few faculty members who really cared about the program and were willing to work very hard  as a team to make
it succeed. Once it succeeded then we moved into maintenance.

\begin{center}
REFERENCES\\
(underlined names refer to students)
\end{center}

\noindent
[1] P.J. Mulvey, E. Dodge and S. Nicholson, {\em Enrollments and degrees report R-151.33},
(American Institute of Physics, April 1997).\\

\noindent
[2] \underline{M. Waldsmith} and J.W. Norbury, {\em Uranium Beam Lifetimes at RHIC and LHC},  5th
Annual Argonne Symposium for undergraduates in Science,  Engineering and
Mathematics, Argonne National Lab, (1994). \\

\noindent 
[3] \underline{B. Soller}, G. Sudhakaran, and M. Jackson, {\em Far-Infrared Laser Stark Spectroscopy of}
$^{13}CH_3 OH$, 5th
Annual Argonne Symposium for Undergraduates in Science, Engineering \& Mathematics, Argonne National
Laboratory, 1994.\\

\noindent 
[4] \underline{P. Valentine}, G. Sudhakaran, and M. Jackson, {\em Far-Infrared Laser Stark Spectroscopy
of } $CH_3 OD$,  5th
Annual Argonne Symposium for Undergraduates in Science, Engineering \& Mathematics, Argonne National
Lab, 1994.\\

\noindent 
[5] \underline{K.J. Cook}, G.R. Sudhakaran, and M. Jackson, {\em Far-Infrared Water Vapor Laser}, 5th
Annual Argonne Symposium for Undergraduates in Science, Engineering and Mathematics, Argonne National
Laboratory, 1994.\\

\noindent 
[6] \underline{J.T. Dobler}, G.R. Sudhakaran, and M. Jackson, {\em Stark Spectroscopy using a Far-Infrared
Laser}, 7th Annual Argonne Symposium for Undergraduates in Science, Engineering and Mathematics, Argonne
National Laboratory, 1996.\\

\noindent 
[7] \underline{E.J. Gansen},  G.R. Sudhakaran and M. Jackson, {\em Far-Infrared Laser Stark Spectroscopy of
$PH_3$}, 7th Annual Argonne Symposium for Undergraduates in Science, Engineering and Mathematics, Argonne
National Laboratory, 1996.\\

\noindent
[8] {\em Regge Trajectories for Mesons} (\underline {M. Pruse}, \JWN) 
5th Annual Argonne Symposium for undergraduates in Science, Engineering
 and Mathematics,
Argonne National Laboratory, 1994.\\

\noindent
[9] {\em Parameterization of spectral distributions for pion and kaon 
production in proton-proton collisions} (\underline {J. Schneider}, \JWN\ and \FAC) 5th
Annual Argonne Symposium for undergraduates in Science, Engineering and Mathematics, Argonne
National Lab, 1994.\\

\noindent
[10] \underline{J.P. Schneider}, J.W. Norbury and F.A. Cucinotta, {\em Parameterization 
of spectral distributions for pion and kaon production from proton-proton collisions},  
Bulletin of the American Astronomical Society {\bf 26},
873 (1994).\\

\noindent 
[11] M. Jackson, \underline{B.J. Soller}, G.R. Sudhakaran, R.M. Lees, and I. Mukhopadhyay, 
{\em Far-Infrared Laser Stark Spectroscopy of} $^{13}CH_3 OH$,
50th International Symposium on Molecular Spectroscopy, Ohio State University, 1995.\\

\noindent 
[12] M. Jackson, G.R. Sudhakaran, and \underline{E.J. Gansen}, {\em Far-Infrared Laser Stark Spectroscopy
of} $PH_3$, 51st International Symposium on Molecular Spectroscopy, Ohio State University, 1996.\\

\noindent 
[13] M. Jackson, G.R. Sudhakaran, and \underline{E.J. Gansen}, {\em Far-Infrared Laser Stark Spectroscopy
of} $^{13}CD_3 OD$, 52nd International Symposium on Molecular Spectroscopy, Ohio State University, 1997.\\

\noindent
[14]  J.W. Norbury and \underline{C.M. Mueller}, {\em Cross Section parameterizations for Cosmic
Ray Nuclei II. Double Nucleon Removal}, Astrophys. J. Suppl. {\bf 90}, 115-117 (1994).\\

\noindent 
[15] \underline{R. Wheeler} and J.W. Norbury, {\em Higher order corrections to Coulomb fission}, 
Phys. Rev. C {\bf 51}, 1566-1567 (1995).\\

\noindent 
[16] G.R. Sudhakaran, \underline{E.K. Coulson}, and M. Jackson, {\em Laser Stark Spectroscopy of}
$^{13}CH_3 F$, Int. J. Infrared Millimeter Waves, {\bf 16}, 1329-1333 (1995).\\

\noindent 
[17] G.R. Sudhakaran, \underline{B.J. Soller}, M. Jackson,  I. Mukhopadhyay,  and R.M. Lees, 
{\em Far-Infrared Laser Stark Spectroscopy of }$CH_3 OH$ and $^{13}CH_3 OH$, Int. J. Infrared Millimeter
Waves, {\bf 16}, 2111-2131 (1995).\\

\noindent 
[18] M. Jackson, G.R. Sudhakaran, and \underline{E.J. Gansen}, {\em Far-Infrared Laser Stark Spectroscopy of
} $^{13}CD_3 OD$, J. Mol. Spectrosc., {\bf 176}, 439-441 (1996).\\

\noindent 
[19] M. Jackson, G.R. Sudhakaran, and \underline{E.J. Gansen}, {\em Far-Infrared Laser Stark Spectroscopy
of} $PH_3 $, J. Mol. Spectrosc., {\bf 181}, 446-451 (1997).\\

\noindent
[20] {\em Parameterized  spectral distributions for meson production in 
proton-proton collisions} (\underline {J.P. Schneider}, \JWN\ and \FAC) \\
\TM\ {\bf 4675} (1995).\\

\noindent
[21] {\em Parameterization of spectral distributions for 
pion and kaon production in proton-proton collisions} 
(\underline {J.P. Schneider}, \JWN\ and
\FAC) \APJS\ {\bf 97} 571-574 (1995).

\end{document}